\documentclass[prd,aps,preprintnumbers,fleqn,showpacs,nofootinbib,superscriptaddress]{revtex4}

\usepackage{amssymb,amsmath,graphicx,bm}

\usepackage[usenames,dvipsnames]{color}
\usepackage[normalem]{ulem} 
\renewcommand\sout{\bgroup \color[rgb]{0.55,0.00,0.99} \ULdepth=-.5ex \ULset}

\def\slash#1{\setbox0=\hbox{$#1$}               
        \dimen0=\wd0                            
        \setbox1=\hbox{/} \dimen1=\wd1          
        \ifdim\dimen0>\dimen1                   
        \rlap{\hbox to \dimen0{\hfil/\hfil}}    
        #1                                      
        \else
        \rlap{\hbox to \dimen1{\hfil$#1$\hfil}} 
        /                                       
        \fi}                                    %

\newcommand{\beqn}{\begin{eqnarray}}
\newcommand{\eeqn}{\end{eqnarray}}

\begin{document}


\title{Suppression of maximal linear gluon polarization in angular asymmetries}

\author{Dani\"el Boer}
\email{d.boer@rug.nl}
\affiliation{ {Van Swinderen Institute for Particle Physics and Gravity, University of Groningen, Nijenborgh 4, 9747 AG Groningen, The Netherlands}}

\author{Piet J. Mulders}
\email{mulders@few.vu.nl}
\affiliation{Nikhef and Department of Physics and Astronomy, VU University
 Amsterdam, De Boelelaan 1081, NL-1081 HV Amsterdam,
The Netherlands}

\author{Jian~Zhou}
\email{jzhou@sdu.edu.cn}
 \affiliation{\normalsize\it School of physics, $\&$ Key Laboratory of
Particle Physics and Particle Irradiation (MOE), Shandong University, Jinan, Shandong 250100,
China}

\author{Ya-jin Zhou}
\email{zhouyj@sdu.edu.cn} \affiliation{\normalsize\it School of physics, $\&$ Key Laboratory of
Particle Physics and Particle Irradiation (MOE), Shandong University, Jinan, Shandong 250100,
China}

\begin{abstract}
We perform a phenomenological analysis of the $\cos 2 \phi $ azimuthal asymmetry in virtual photon
plus jet production induced by the linear polarization of gluons in unpolarized $pA$ collisions.
Although the linearly polarized gluon distribution becomes maximal at small $x$, TMD evolution
leads to a Sudakov suppression of the asymmetry with increasing invariant mass of the
$\gamma^*$-jet pair. Employing a small-$x$ model input distribution, the asymmetry is found to be
strongly suppressed under TMD evolution, but still remains sufficiently large to be measurable in
the typical kinematical region accessible at RHIC or LHC at moderate photon virtuality, whereas it
is expected to be negligible in $Z/W$-jet pair production at LHC. We point out the optimal
kinematics for RHIC and LHC studies, in order to expedite the first experimental studies of the
linearly polarized gluon distribution through this process. We further argue that this is a
particularly clean process to test the $k_t$-resummation formalism in the small-$x$ regime.
\end{abstract}

\date{\today}

\maketitle

\section{Introduction}
The linearly polarized gluon distribution  has received growing attention from both the small $x$
physics and the spin physics community in recent years. It is the only polarization dependent gluon
transverse momentum dependent distribution (TMD) inside an {\it unpolarized} nucleon or nucleus at
leading power. The linearly polarized gluon distribution denoted as $h_1^{\perp g}$ was first
introduced in Ref.~\cite{Mulders:2000sh}, and later was discussed in a model in
Ref.~\cite{Meissner:2007rx}. From a different point of view, it was also considered in the context
of $k_t$ resummation~\cite{Nadolsky:2007ba,Mantry:2009qz,Catani:2010pd}. The linearly polarized
gluon distribution so far has not yet been studied experimentally. It has been proposed to probe
$h_1^{\perp g}$ by measuring $\cos 2\phi$ azimuthal asymmetry for two particle production in
various processes ~\cite{Boer:2009nc,Boer:2010zf,Qiu:2011ai,Pisano:2013cya}. In all these cases,
the maximal asymmetries allowed by the positivity bound constraint for the linearly polarized gluon
distribution turn out to be sizable. These findings are quite promising concerning a future
extraction of $h^{\perp g}_{1}$ at RHIC, LHC, or a future Electron-Ion Collider (EIC). It has also
been noted that $h^{\perp g}_{1}$ affects the angular independent transverse momentum distribution
of scalar or pseudoscalar particles, such as the Higgs
boson~\cite{Boer:2011kf,Sun:2011iw,Wang:2012xs,Boer:2013fca,Catani:2013tia,Boer:2014tka,Echevarria:2015uaa}
or  charmonium and bottomonium
states~\cite{Boer:2012bt,Ma:2012hh,Dunnen:2014eta,Zhang:2014vmh,Ma:2015vpt} at LHC.

Theoretical studies of the linearly polarized gluon distribution indicate that it can be quite
sizable compared to the unpolarized gluon distribution. In the DGLAP formalism its small-$x$
asymptotic behavior is the same as the unpolarized gluon distribution, implying that grows equally
rapidly towards small $x$. In the McLerran-Venugopalan (MV) model~\cite{McLerran:1993ni} that
describes gluon saturation at small $x$, it has been found~\cite{Metz:2011wb} that the linearly
polarized gluon distribution inside a large nucleus (or nucleon) reaches its maximal value allowed
by the positivity constraint for transverse momenta above the saturation scale $Q_s$. At low
transverse momentum, the gauge link structure of the gluon TMDs becomes relevant, which is
intimately connected to the process considered. For the Weizs\"{a}cker-Williams (WW) type $h^{\perp
g}_{1}$ containing a staple-like gauge link, the gluon linear polarization is suppressed in the
dense medium region, while for the dipole type $h^{\perp g}_{1}$ where the gauge link is a closed
Wilson loop, it saturates the positivity bound~\cite{Metz:2011wb}. It is further shown in
Ref.~\cite{Dominguez:2011br} that the small-$x$ evolution of the dipole type polarized gluon TMD
and the normal unpolarized gluon TMD is governed by the same nonlinear evolution (BK) equation. In
contrast, the WW type $h^{\perp g}_{1}$ does not rise as rapidly as the WW type unpolarized gluon
TMD does towards small $x$ region. The phenomenological implications of the large gluon linear
polarization inside a large nucleus have been explored in
Refs.~\cite{Metz:2011wb,Dominguez:2011br,Schafer:2012yx,Liou:2012xy,Akcakaya:2012si,Dumitru:2015gaa,Dumitru:2016jku}.
The process under consideration in the present paper probes the dipole type TMDs, for which the
linear gluon polarization is expected to become maximal at small $x$.

To reliably extract small-$x$ gluon TMDs in high energy scattering, it is of great importance to
first establish TMD factorization. As a leading power approximation, TMD factorization is usually
expected to hold at moderate to large $x$~\cite{Collins:1981uk}. At small $x$, however, higher
twist contributions are equally important as the leading twist contribution because of the high
gluon density. In order to arrive at an effective TMD factorization at small $x$, one first has to
calculate the complete hard scattering cross section in the Color Glass Condensate (CGC) framework,
 which is expressed as the convolution of hard part and Wilson lines.
 The next step is to extrapolate the full CGC result to the correlation limit
where the gluon transverse momentum is much smaller than the hard scale(s) in the process.
 One can then justify the use of TMD factorization at small $x$, by reducing the complete CGC result to
 the cross section formula derived in TMD factorization. Such an effective TMD factorization has
 been established in various high energy scattering processes in $eA$ and $pA$ collisions at large $N_c$~\cite{Dominguez:2010xd,Dominguez:2011wm}.
 This has been extended to finite $N_c$ and polarization dependent cases for heavy quark
 production in $pA$ collisions in Ref.~\cite{Akcakaya:2012si}, as well as to other channels for two
 particle production in $pA$ collisions~\cite{Kotko:2015ura,vanHameren:2016ftb}.
 At this point, we emphasize that with the help of the effective TMD factorization that is derived from
 the CGC approach, the phenomenological analysis of the
 relevant physical observables can be greatly simplified in a certain kinematical region.

 In this paper, we study the $\cos 2\phi$ azimuthal asymmetry for virtual photon-jet production in the
 forward region in $pA$ collisions, i.e., $p+A\rightarrow \gamma^*+q+X$. Here, the azimuthal
 angle $\phi$ refers to the angle between the transverse momentum of the $\gamma^*$-$q$ pair ($k_\perp$) and that of
 the virtual photon or the jet ($P_\perp$) in the back-to-back correlation limit ($k_\perp\ll P_\perp$).
 From a theoretical point of  view, this is the cleanest and simplest process to access the dipole type linearly polarized
 gluon distribution. The back-to-back correlation limit is essential here, because only in this limit one
 finds a full match between the CGC result and the effective TMD factorization~\cite{Metz:2011wb,Dominguez:2011br}.
We will present some technical details in the next section showing how to extrapolate the CGC
result to the correlation limit. However, this is not yet the complete story. Due to the fact that
there exists two well separated scales $k_\perp^2$ and $P_\perp^2$ in the process under
consideration, improving the perturbative calculation in a systemic way requires to resum to all
orders the large logarithms $\alpha_s \ln^2 P_\perp^2/k_\perp^2$ that show up in higher order
corrections. It has been shown in Ref.~\cite{Mueller:2012uf,Mueller:2013wwa} that such resummation
can be done consistently within the CSS formalism in the saturation regime. As a result, the
standard double logarithm Sudakov form factor emerges in the effective TMD factorization formula,
which leads to the suppression of the asymmetry as shown below. In addition, more recent
work~\cite{Zhou:2016tfe} indicates that the single logarithm $\alpha_s \ln P_\perp^2/k_\perp^2$
also can be consistently resummed in the small-$x$ formalism at least in the dilute limit. Both
 the double and single logarithms are included in our phenomenological analysis of the azimuthal
asymmetry following the standard CSS formalism. We note that another formulation of the small-$x$
Sudakov resummation exists in the literature~\cite{Balitsky:2015qba}, but relating the two
approaches will not be attempted here.

For completeness it should be mentioned that the linearly polarized gluon distribution is not the
only spin-dependent gluon distribution that is relevant at small $x$. It was found in a sequence of
papers~\cite{Schafer:2013opa,Zhou:2013gsa,Boer:2015pni,Szymanowski:2016mbq} that the Sivers gluon
distribution of the dipole type is not suppressed by a full power of $x$ with respect to the
unpolarized gluon TMD towards small $x$, whereas the WW type one is. Furthermore, it has been shown
in Refs.~\cite{Hatta:2016dxp,Hagiwara:2016kam,Zhou:2016rnt,Iancu:2017fzn} that the polarization
dependent five-dimensional {\it generalized} TMD inside a large nucleus could be sizable. In
addition, several spin-dependent gluon TMDs inside a spin-1 target that could persist in the
small-$x$ limit are identified in Ref.~\cite{Boer:2016xqr}. It would be very interesting to test
these theoretical expectations at RHIC, LHC or a future EIC.

The paper is structured as follows. In the next section, we  describe the general theoretical
framework, including justifying the use of effective TMD factorization from a CGC expression, the
discussion of the associated factorization properties, and incorporating  the Sudakov suppression
effect. In Sec.~III, we present the numerical results for the azimuthal asymmetry in the various
kinematical regions potentially accessible at RHIC and LHC. A summary of our findings and
conclusions is presented in Sec.~IV.

\section{Theoretical setup}
The virtual photon-jet production in the forward region in $pA$ collision is dominated by the
partonic process,
\begin{eqnarray}
q(x_p \bar P)+g(x P+k_\perp) \rightarrow \gamma^*(p_1)+q(p_2),
\end{eqnarray}
where $x P+k_\perp$ is understood as the total momentum transfer through multiple gluon exchange
when a quark from proton scattering off the gluon background inside a large nucleus. Typically one
computes the cross section for this process using a hybrid approach~\cite{Gelis:2002ki} in which
the dense target nucleus is treated as a CGC, while on the side of dilute projectile proton one
uses the ordinary integrated parton distribution functions (PDFs). Although a general proof of this
method is still lacking, from a practical point of view such a hybrid approach is very useful and
we will employ it here. Problematic contributions causing a violation of (generalized) TMD
factorization~\cite{Rogers:2010dm} are absent in this formalism. This is because one still can
employ the Ward identity argument to decouple longitudinal gluon attachments from the proton side
like in collinear factorization. Put differently, the factorization breaking terms are suppressed
by powers of $\Lambda_\text{QCD}^2/Q_s^2$ in the semi-hard region where the imbalance transverse
momentum of the virtual photon-jet system is of the order of the saturation scale $Q_s$. For more
detailed arguments why the factorization breaking effects can be avoided in the semi-hard region,
we refer readers to Refs.~\cite{Akcakaya:2012si,Schafer:2014zea,Schafer:2014xpa,Zhou:2015ima}.

It is straightforward to obtain the production amplitude in this hybrid
approach~\cite{Gelis:2002ki},
\begin{eqnarray}
{\cal M}=H(k_\perp) \left [ U(k_\perp)-(2\pi)^2 \delta^2(k_\perp) \right ],
\end{eqnarray}
with $H(k_\perp)$ given by,
\begin{eqnarray}
H(k_\perp)=\bar u(p_2) \left [ (ie)\varepsilon\!\!\!/ i \frac{x_p \bar P \!\!\!\!/ +x
 P\!\!\!\!/+k_\perp\!\!\!\!\!\!/}{(x_p \bar P+xP+k_\perp)^2+i\epsilon} p\!\!\!/
 +p\!\!\!/i  \frac{x_p \bar P \!\!\!\!/ - p_1\!\!\!\!\!/}{(x_p \bar P- p_1)^2+i\epsilon}(ie)\varepsilon\!\!\!/
 \right ]u(x_p \bar P),
\end{eqnarray}
where $\varepsilon^\mu$ is the polarization vector of the produced virtual photon. The Wilson line
$U(k_\perp)$ is defined as,
\begin{eqnarray}
U(k_\perp)=\int d^2 x_\perp e^{ik_\perp \cdot x_\perp} {\cal P} e^{ig\int_{-\infty}^{+\infty} dx^-
A^+(x^-,x_\perp) \cdot t}.
\end{eqnarray}
The above expresses that in the small $x$ limit, the incoming quark from the proton interacts
coherently with the nucleus as a whole. These interactions are summarized into Wilson lines which
stretch from minus infinity to plus infinity. The next step is to extrapolate this CGC result to
the correlation limit either in  coordinate
space~\cite{Dominguez:2010xd,Dominguez:2011wm,Dominguez:2011br} or in momentum
space~\cite{Metz:2011wb,Akcakaya:2012si}. Here we choose the latter and introduce the two momenta $
P_\perp=( p_{1\perp}- p_{2\perp})/2$ and $k_\perp=p_{1\perp}+p_{2\perp}$. In the correlation limit,
one has $|P_\perp|\simeq |p_{1\perp}|\simeq |p_{2\perp}|\gg |k_\perp|=|p_{1\perp}+p_{2\perp}|$. The
azimuthal angle between $k_\perp$ and $P_\perp$ will be denoted by $\phi$. In the correlation
limit, the additional hard scale $P_\perp^2$ ensures that the hard scattering takes place locally
where only a single gluon exchange from the nucleus takes part in the interaction. Multiple
exchanges are power suppressed. This corresponds to a Taylor expansion of the hard part,
\begin{eqnarray}
H(k_\perp)=H(k_\perp=0)+ \left. \frac{\partial H(k_\perp)}{\partial k_\perp^i}\right|_{k_\perp=0}k_\perp^i+... ,
\end{eqnarray}
where the first term does not contribute and the neglected terms are suppressed by powers of
$|k_\perp|/|p_{2\perp}|$. The cross section is then calculated by squaring the amplitude,
\begin{eqnarray}
d \sigma \propto  \left.\frac{\partial H(k_\perp)}{\partial k_\perp^i}\right|_{k_\perp=0} \left.\frac{\partial
H^*(k_\perp)}{\partial k_\perp^j}\right|_{k_\perp=0} \int d^2 x_\perp d^2 y_\perp k_\perp^i k_\perp^j \
e^{i(x_\perp-y_\perp)\cdot k_\perp} \left\langle {\rm Tr} \left [ U^\dag(y_\perp)U(x_\perp) \right ]
\right\rangle.
\end{eqnarray}
Using the formula,
\begin{eqnarray}
\partial^i U(x_\perp)=-i g \int_{-\infty}^\infty dx^- U[-\infty,x^-,x_\perp] \partial^iA^+(x^-,x_\perp) U[x^-,+\infty,x_\perp] \ ,
\end{eqnarray}
one can identify the soft part as the gluon TMD matrix element after partial integration.
 Effectively it means that in the correlation limit the incoming quark no longer interacts
coherently with the nucleus throughout the process, but that the leading power effect of the
interactions with the classical gluonic state are restricted to before and after the hard
scattering. The net effect is an $SU(3)$ color rotation of the incoming quark and the outgoing
quark-virtual photon system that is encoded in the gauge links of the TMD. We parameterize the
gluon TMD matrix element as,
\begin{eqnarray}
\int d^2 x_\perp d^2 y_\perp  \ e^{i(x_\perp-y_\perp)\cdot k_\perp} \langle {\rm Tr} \left [
\partial^i U^\dag(y_\perp)\partial^i U(x_\perp) \right ] \rangle
 \propto \frac{\delta_{\perp}^{ij}}{2} \, x f_1^g(x, k_\perp) + \bigg(\frac{ k_\perp^i
k_\perp^j}{k_\perp^2} - \frac{1}{2} \delta_{\perp}^{ij} \bigg) x h^{\perp g}_{1}(x, k_\perp) \, ,
\end{eqnarray}
where $ f_1^g(x, k_\perp)$ is the regular unpolarized gluon TMD. Note that the convention for
$h_1^{\perp g}$ used here differs from~\cite{Mulders:2000sh} by a factor $k^2_\perp/2M_p^2$, such
that positivity bound reads $| h_1^{\perp g}(x, k_\perp)| \leq f_1^g(x, k_\perp)$.  By contracting
the hard part with the tensor $\frac{\delta_{\perp}^{ij}}{2}$, one obtains the azimuthal
independent cross section, while contracting the hard part with the tensor $\bigg(\frac{ k_\perp^i
k_\perp^j}{k_\perp^2} - \frac{1}{2} \delta_{\perp}^{ij} \bigg)$ produces a $\cos 2\phi$ modulation.
Collecting all these ingredients, we eventually arrive at the cross section
formula~\cite{Metz:2011wb},
\begin{eqnarray}
\frac{d\sigma^{pA\to \gamma^*qX}}{dP.S}&=&\sum_q x_p  f_1^q(x_p)   \Big{\{} x f^g_{1}(x,k_{\perp})
H_{\text{Born}}
 + \text{cos}(2\phi)\, x h^{\perp g}_{1}(x,k_{\perp}) H_{\text{Born}}^{\text{cos}(2\phi)} \Big{\}},
 \label{eqn:D cross section}
\end{eqnarray}
where $f_1^q(x_p)$ is the quark collinear PDF of the proton and the hard coefficients are given by,
\begin{eqnarray}
\nonumber H_{\text{Born}} &=& \frac{\alpha_s \alpha_{em} e^2_q (1-z) z^2}{N_c}
 \left[  \frac{1+(1-z)^2}{\left(P_{\perp}^2+(1-z)Q^2\right)^2}
 -\frac{2 Q^2 P^2_{\perp} z^2 (1-z)}{\left(P_{\perp}^2+(1-z)Q^2\right)^4} \right ]
 \\ \nonumber
H_{\text{Born}}^{\text{cos}(2\phi)}&=& \frac{\alpha_s \alpha_{em} e^2_q (1-z) z^2}{N_c}  \frac{-2 Q^2
P^2_{\perp} z^2 (1-z)}{\left(P_{\perp}^2+(1-z)Q^2\right)^4}
\end{eqnarray}
which is in full agreement with that derived from TMD factorization~\cite{Metz:2011wb}.

In the above formula, the phase space factor is defined as $d P.S=dy_q dy_{\gamma^*} d^2P_\perp
d^2 k_\perp$, where $y_q$ and $ y_{\gamma^*}$ are the rapidities of the produced quark and the
virtual photon respectively. $Q^2$ and $z$ are the virtual photon invariant mass and  the
longitudinal momentum fraction of the incoming quark carried by the virtual photon, respectively.

The next step is to resum the large logarithm $\ln P_\perp^2/k_\perp^2$ that arises from higher
order corrections. An explicit one-loop calculation of the scalar particle production has shown
that the $k_t$ resummation can be consistently done using the standard CSS formalism within the CGC
effective theory framework~\cite{Mueller:2012uf}. It has been further demonstrated in
Ref.~\cite{Mueller:2013wwa} that the double leading logarithm terms can be resummed into a Sudakov
form factor for photon-jet production in the unpolarized case. More evidence that the conventional
$k_t$ resummation procedure in general is compatible with the small-$x$ formalism has been found in
Ref.~\cite{Zhou:2016tfe}, implying that both the double leading logarithm and single leading
logarithm can be resummed into the Sudakov form factor for the unpolarized case as well as the
polarized case by means of the Collins-Soper evolution equation.

To facilitate resumming the large $k_t$ logarithm, one should Fourier transform the cross section
formula to $b_\perp$ space and insert the Sudakov form factor following the standard CSS formalism,
leading to
\begin{eqnarray}\nonumber
\frac{d\sigma^{pA\to \gamma^*qX}}{dP.S}&=&\sum_q \int d^2 b_{\perp}~ e^{i k_{\perp}\cdot b_{\perp}}
 ~x_p f_1^q(x_p,\mu_b^2)e^{-S\left(\mu_{b}^2,
P_\perp^2 \right)}
 \\&\times& \Big{\{} x f^g_{1}(x,b_{\perp}^2,\mu_b^2)~ H_{\text{Born}}
 + x h^{\perp g}_{1}(x,b_{\perp}^2,\mu_b^2)
\left [ 2(\hat{b}_{\perp} \cdot \hat{P}_{\perp})^2-1 \right ] H_{\text{Born}}^{\text{cos}(2\phi)} \Big{\}},
\end{eqnarray}
where $\mu_b=2 e^{-\gamma_E}/|b_{\perp}| $, $\hat b_\perp=b_\perp/|b_\perp|$ and $\hat
P_\perp=P_\perp/|P_\perp|$ are unit vectors. The unpolarized and polarized gluon TMDs in $b_\perp$
space are given by,
\begin{eqnarray}
  x f^g_{1}(x,b_{\perp}^2,\mu_b^2) &=& \int \frac{d^2 {k}_{\perp}}{(2 \pi)^2}
  ~e^{i {k}_{\perp}\cdot {b}_{\perp}}
 x f_{1}^g(x,k_{\perp},\mu_b^2),
 \label{eqn:xf1DPk}
\\
 x h^{\perp g}_{1}(x,b_{\perp}^2,\mu_b^2) &=& -\int \frac{d|k_{\perp}|}{2\pi}
J_2(|b_{\perp}||k_{\perp}|) ~ x h^{\perp g}_{1}(x,k_{\perp},\mu_b^2), \label{eqn:xh1DPk}
\end{eqnarray}
where the standard $\zeta$ parameter is chosen identical to the renormalization scale $\mu_b$ and
not shown here. At tree level the Sudakov factor is zero, in which case one recovers
Eq.~(\ref{eqn:D cross section}) after Fourier transforming back to $k_\perp$ space. At one-loop
order, the perturbative Sudakov form factor (valid for sufficiently small $b_\perp$) takes the
form,
\begin{eqnarray}
\label{eqn:Sudakov} S(\mu_b^2, P_\perp^2)= \int_{\mu_b}^{|P_\perp|} \frac{d\mu}{\mu} \alpha_s(\mu)
\left(\frac{C_F+C_A}{\pi} \ln \frac{P_\perp^2}{\mu^2} -\frac{C_F}{\pi} \frac{3}{2}
-\frac{C_A}{\pi} \frac{11-2 n_f/C_A}{6} \right ),
\end{eqnarray}
where the $C_F$ part receives a contribution from a gluon radiated off the quark line, while the
$C_A$ part is generated from the Collins-Soper type small-$x$ gluon TMD evolution. We will discuss
the nonperturbative Sudakov factor in the next section.

To avoid having to deal with a three scale problem,
we restrict to the kinematical region where $Q^2$ is of the order of $P_\perp^2$.  This happens
to be the optimal region to probe $ h^{\perp g}_{1}$ as suggested by our numerical estimation.

\section{numerical results}
To evolve the gluon TMDs to a higher scale, one has to first determine the gluon TMDs at an initial
scale. It is common to compute the  gluon distributions in the MV model and use them as the initial
condition for small $x$ evolution. At RHIC energy, the typical longitudinal momentum fraction
carried by gluon probed in the process under consideration is around $x \sim 0.01$, which we
consider to be sufficiently small to apply a small-$x$ input distribution. Because of the limited
$x$ range probed at RHIC, we do not include small-$x$ evolution. At LHC this may become relevant
though.

We will use the MV model results as
the initial input for the Collins-Soper type evolution. In the MV model, the dipole type
unpolarized and linearly polarized gluon TMDs are identical~\cite{Metz:2011wb},
\begin{eqnarray}
 x h^{\perp g}_{1}(x,k_{\perp})=xf^g_{1}(x,k_{\perp}) = \frac{k^2_{\perp} N_c }{2 \pi^2 \alpha_s} S_{\perp} \int \frac{ d^2
{b}_{\perp}}{(2 \pi)^2} ~e^{-i {k}_{\perp} \cdot {b}_{\perp} } e^{-\frac{b_{\perp}^2 Q_s^2}{4}}
\end{eqnarray}
where $S_{\perp}$ denotes the transverse area of a large nucleus. To facilitate numerical
estimation, we reexpress it as~\cite{Mueller:1999wm} ,
\begin{eqnarray}
x f^g_{1}(x,k_{\perp}) =\frac{ k_{\perp}^2 A~ x f_{1,p}^g(x)}{Q^2_s}\int \frac{ d^2 {b}_{\perp}}{(2
\pi)^2} ~e^{-i {k}_{\perp} \cdot {b}_{\perp} } e^{-\frac{b_{\perp}^2 Q_s^2}{4}}
 \label{eqn:ungluonGBW}
\end{eqnarray}
with $f_{1,p}^g(x)$ being the standard gluon PDF in a nucleon, for which we will employ the MSTW
2008 LO PDF set. Substituting Eq.~(\ref{eqn:ungluonGBW}) to Eq.~(\ref{eqn:xf1DPk}) and
Eq.~(\ref{eqn:xh1DPk}), one obtains,
\begin{eqnarray}
x f^g_{1}(x,b_{\perp}^2,\mu^2\equiv Q_s^2) &=&A x f_{1,p}^g(x,\mu^2\equiv Q_s^2)\frac{1}{(2 \pi)^2}
\left (1 -\frac{Q_s^2 b_{\perp}^2}{4}\right ) e^{-\frac{Q_s^2 b^2_{\perp}}{4}}
\\
x h^{\perp g}_{1}(x,b_{\perp}^2,\mu^2\equiv Q_s^2) &=& - A x f_{1,p}^g(x,\mu^2\equiv
Q_s^2)\frac{1}{(2 \pi)^2} \frac{Q_s^2 b_{\perp}^2}{4} e^{-\frac{Q_s^2 b^2_{\perp}}{4}}
\end{eqnarray}
In arriving at the above formulas, we have neglected the dependence of $Q_s^2$ on $b_\perp^2$ that
is usually considered as a good approximation at low $k_\perp$, although it leads to an incorrect
perturbative tail at high transverse momentum~\cite{Dominguez:2011wm}. In the current case,  the
correct power law tail automatically develops after taking into account TMD evolution. It should be
also pointed out that the renormalization scale and the parameter $\zeta$ have been chosen to be
$Q_s^2$, at which scale the MV model expressions are assumed to hold.

Before computing the asymmetry, it would be interesting to first investigate how the linearly gluon
polarization is affected by TMD evolution. By solving the Collins-Soper equation, the unpolarized
and polarized gluon TMDs at the scales  $P_\perp^2$  read,
\begin{eqnarray}
 x f_{1}^g (x,k_{\perp},\mu^2=P_\perp^2)
 &=& \int d |b_{\perp}| ~|b_{\perp}| ~2 \pi J_0(|k_{\perp}|| b_{\perp}|)~ e^{-S_A(\mu_b^2,P_\perp^2)}~
 x f_{1}^g(x,b_{\perp }^2,\mu_b^2), \\
  x h_{1}^{\perp g} (x,k_{\perp},\mu^2=P_\perp^2)&=&-\int d |b_{\perp}| ~|b_{\perp}| ~2 \pi
  J_2(|k_{\perp}||b_{\perp}|)~ e^{-S_A(\mu_b^2,P_\perp^2)}~ x h_{1}^{\perp g}(x,b_{\perp }^2,\mu_b^2),
\end{eqnarray}
and similar expressions holds at the scale $Q_s^2$ after replacing $P_\perp^2$ by $Q_s^2$. For the
expressions at $Q_s^2$ we will then use the MV model expressions.
The gluonic part of the perturbative Sudakov factor $S_A(\mu_b^2,P_\perp^2)$ at one-loop order
takes the form,
\begin{eqnarray}
 S_A(\mu_b^2,P_\perp^2) =
\frac{C_A}{2\pi} \int^{P_\perp^2}_{\mu_b^2} \frac{d\mu^2}{\mu^2} \alpha_s(\mu) \left[ \ln
\frac{P_\perp^2}{\mu^2} - \frac{11-2n_f/C_A}{6} \right ],
\end{eqnarray}
Using the above relations, it follows that
\begin{eqnarray}
  x f_{1}^g (x,k_{\perp},\mu^2=P_\perp^2)
 &=& \int d |b_{\perp}| ~|b_{\perp}| ~2 \pi J_0(|k_{\perp}|| b_{\perp}|)~ e^{-S_A(\mu_b^2,P_\perp^2)+S_A(\mu_b^2,Q_s^2)}~
 x f_{1}^g(x,b_{\perp }^2,\mu^2=Q_s^2), \\
  x h_{1}^{\perp g} (x,k_{\perp},\mu^2=P_\perp^2)&=&-\int d |b_{\perp}| ~|b_{\perp}| ~2 \pi
  J_2(|k_{\perp}||b_{\perp}|)~ e^{-S_A(\mu_b^2,P_\perp^2)+S_A(\mu_b^2,Q_s^2)}~ x h_{1}^{\perp g}(x,b_{\perp }^2,\mu^2=Q_s^2),
\end{eqnarray}
where the Sudakov factor $-S_A(\mu_b^2,P_\perp^2)+S_A(\mu_b^2,Q_s^2)$ is the same as that used for
TMD evolution from a fixed scale~\cite{Boer:2013zca}. Note that this is not simply
$S_A(P_\perp^2,Q_s^2)$, due to the double log nature of the expressions.

The above expressions for the Sudakov factor are valid in the perturbative region of small
$b_\perp$. Since we are mostly interested in the semi-hard region $|k_\perp| \sim Q_s$ where the
contributions from large $b_\perp$ could be important when performing Fourier transform, following
the standard treatment, we introduce a non-perturbative Sudakov factor, for both $\mu^2=P_\perp^2$
and $\mu^2=Q_s^2$:
\begin{eqnarray}
-S_A(\mu_b^2,\mu^2) \rightarrow -S_A(\mu_{b*}^2,\mu^2) -S^{NP}(b_\perp^2,\mu^2)
\end{eqnarray}
where $\mu_{b*}^2$ is defined as $\mu_{b*}^2=4e^{-2\gamma_E}/b_{\perp*}^2$, with $b_{\perp*}$
given by
\begin{eqnarray}
b_{\perp *}=\frac{b_{\perp}}{\sqrt{1+b_{\perp}^2/b^2_{\max}}}\leq b_{\max}
\end{eqnarray}
and the parametrization for the non-perturbative Sudakov factor will be taken (based
on~\cite{Aybat:2011zv}) as
\begin{equation}
S_{A,q}^{NP}(b_{\perp}^2,\mu^2)= S_{A,g}^{NP}(b_{\perp}^2,\mu^2) \frac{C_F}{C_A} =
\frac{1}{2}\left( g_1+g_2 \ln \frac{\mu}{2 Q_0} + 2 g_1 g_3 \ln \frac{10 x x_0}{x_0+x} \right)
b_{\perp}^2,
\end{equation}
with $b_{\max}=1.5\, \text{GeV}^{-1},~ g_1=0.201\, \text{GeV}^2, ~ g_2=0.184\,
\text{GeV}^2,~g_3=-0.129,~x_0=0.009,~Q_0=1.6\, \text{GeV}$. To smoothly match to large transverse
momentum region, we also regulate the very small $b_\perp$ behavior of the Sudakov factor by the
replacement of $\mu_b$ by~\cite{Boer:2014tka,Boer:2015uqa,Collins:2016hqq}
\begin{equation}
\mu_b^\prime(\mu^2)= \frac{1}{\sqrt{b^2_{\perp}/(4e^{-2\gamma_E})  +1/\mu^2}}.
\end{equation}
In our numerical estimation,  we used the one-loop running coupling constant $\alpha_s$, with
$n_f=3$ and $\Lambda_{\text{QCD}} = 216~ \text{MeV}$. The saturation scale is further fixed using
the GBW model~\cite{GolecBiernat:1998js},
 \begin{eqnarray}
 Q^2_s(x)=(1\ \text{GeV})^2 A^{1/3}
\left(\frac{x_0}{x}\right)^{0.3}~~~\text{ with}~~ x_0=3\times 10^{-4}
\end{eqnarray}
where the atomic number $A$ is chosen to be $A=197$ for RHIC, but $A=208$ for LHC hardly makes a
difference. It results in $Q_s^2 \sim 2 \, \text{GeV}^2$ for RHIC in the kinematical regions under consideration.

\begin{figure}[htbp]
\begin{center}
\includegraphics[angle=0,scale=1.1]{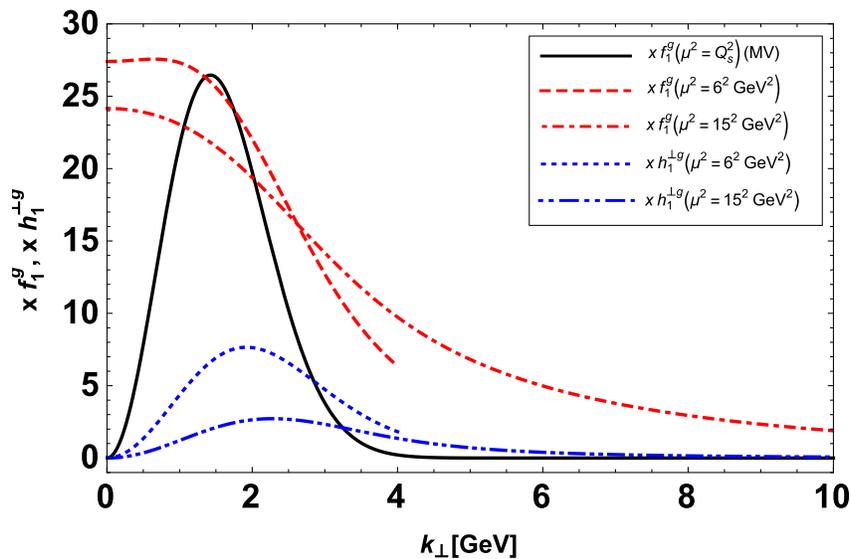}
\caption{The unpolarized and linearly polarized gluon TMDs as function of $k_{\perp}$ at different
scales, at $x$=0.01, using the MV model as input.} \label{figs:gluonTMDs}
\end{center}
\end{figure}
\begin{figure}[hbtp]
\begin{center}
\includegraphics[angle=0,scale=1.1]{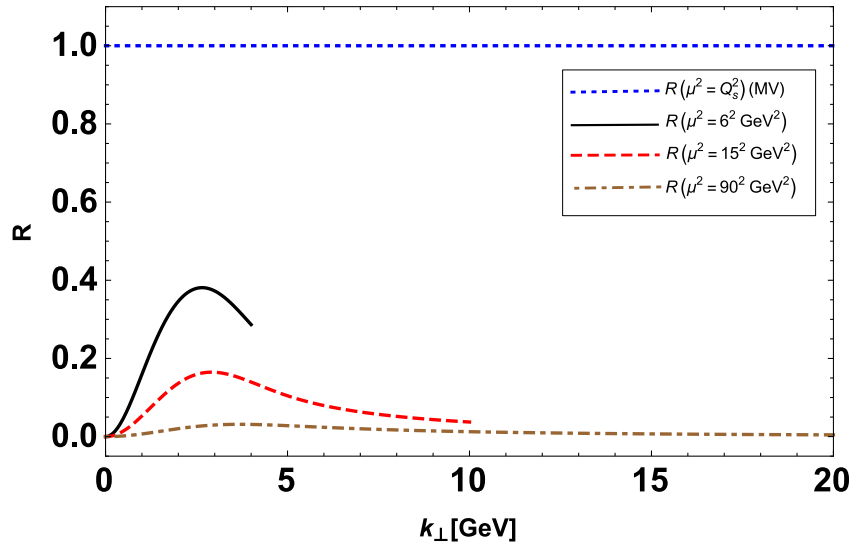}
\caption{The ratio $R=h_1^{\perp g}/f_1^g$ as function of $k_{\perp}$, at $x=0.01$ for $\mu=6$, 15 and 90 GeV.} \label{figs:ratio of gluon TMDs}
\end{center}
\end{figure}
With these ingredients, we are ready to perform the numerical study of the evolved gluon TMDs
including both unpolarized and linearly polarized gluons. We evolve the gluon TMDs from the
saturation scale where the MV results are used as the initial input up to the scales $\mu^2=(6 \,
\text{GeV})^2$ and $\mu^2=(15\  \text{GeV})^2$. As can be seen from Fig.~1, the shape of the
unpolarized gluon TMD  at low $k_\perp$ is significantly changed by evolution. One further observes
that the linearly polarized gluon distribution evolves very fast and is suppressed with increasing
energy. Since the azimuthal asymmetry induced by the linearly polarized gluon distribution is
proportional to the ratio $R=h_1^{\perp g}/f_1^g$, it is instructive to plot this ratio as function
of $k_\perp$ at different scales in Fig.~2. The dotted blue line represents the ratio computed from
the MV model, which is identical to 1 at any value of $k_\perp$. That simple relation between the
unpolarized gluon TMD and the linearly polarized gluon TMD still holds after taking into account
small $x$ evolution~\cite{Dominguez:2011br}, whereas it significantly deviates from it after energy
evolution as shown in Fig.~2. The ratio initially grows with increasing $k_\perp$ until it reaches
a maximal value at a transverse momentum of about two times the saturation scale and then decreases
at high transverse momentum. In the typical kinematical region accessible at RHIC, the maximal
value of the ratio is slightly less than $0.4$. We also plot a curve  for the ratio at the scale
$\mu^2=(90\, \text{GeV})^2$ because it is relevant for studying $\cos 2\phi$ asymmetry for $Z$-jet
pair production in $pp$ or $pA$ collisions at LHC. Judging from this curve for $R$, we conclude
that it is rather challenging to measure the mentioned azimuthal asymmetry at LHC in this process
(and likely in $W$-jet pair production as well). Needless to say, $\gamma^*$-jet pair production at
lower $\gamma^*$ virtuality should be feasible at LHC. There the $x$ values typically are much
smaller, but $Q_s^2$ is only about a factor of 2 larger than at RHIC, requiring $P_\perp$ to
be larger than say 10 GeV, allowing it to still be selected far below $M_Z$. We will also show some
results for LHC below.

\begin{figure}[hbtp]
\begin{center}
\includegraphics[angle=0,scale=1.1]{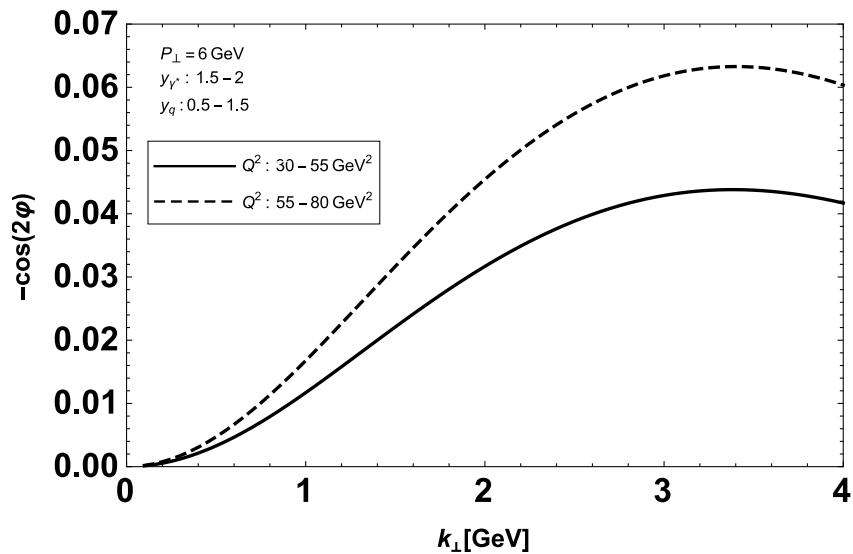}
\caption{Estimates of the azimuthal asymmetry as function of $k_{\perp}$, calculated at
$P_\perp=6 \ \text{GeV}$, for the different $Q^2$ ranges at the center mass energy $\sqrt{s}$=200 GeV. The
quark and virtual photon rapidities are integrated over the regions $y_q$[1.5, 2] and
$y_{\gamma^*}$[0.5, 1.5].} \label{figs:v2 as kperp}
\end{center}
\end{figure}

The numerical results for the computed azimuthal asymmetry in the different kinematical regions are
presented in Figs.~3-5. Here, the azimuthal asymmetry, i.e., the average value of
$\text{cos}(2\phi)$, is defined as,
\begin{eqnarray}
 \langle \text{cos}(2\phi)  \rangle =
 \frac{\int \frac{d\sigma}{d P.S} d\phi ~\text{cos}(2\phi)}{\int \frac{d\sigma}{d P.S} d\phi}.
\end{eqnarray}
As explained in the above, to avoid dealing with the three scales problem, we choose $Q^2$ to be
the order of $P_\perp^2$ which sets the hard scale $\mu^2=P_\perp^2$ when performing energy
evolution. We found that the most optimistic rapidity ranges for measuring this azimuthal asymmetry
at RHIC energy ($\sqrt{s}$=200 GeV) are $y_{\gamma^*}$[1.5, 2], $y_q$[0.5, 1.5]. In these rapidity
ranges, Fig.~3 shows the asymmetry as function of $k_\perp$ for two different $Q^2$ ranges at
$P_\perp=6 \ \text{GeV}$. The asymmetry reach a maximal value of 6\% for $Q^2$[55, 80] around $k_\perp=3 \
\text{GeV}$. Note that the corresponding longitudinal momentum fraction of the gluon $x$ is in the region
[0.008, 0.03] where the MV model results are only borderline justified at best. At LHC the
situation is better in this respect, but here we just aim to illustrate the effect of TMD evolution
on an observable that in principle probes the linear polarization distribution that becomes maximal
at small $x$. Sudakov suppression shows that the observable asymmetry is far from maximal.

For the same rapidity regions, we also plot the asymmetry as function of $P_\perp$ with $Q^2$
chosen to be $Q^2= P_\perp^2$ and $Q^2= \frac{1}{2}P_\perp^2$, with $k_\perp=2.5\text{GeV}$. 
One sees that there is  a relatively mild dependence of
the asymmetry on $P_\perp$.  Fig.~5 shows that the asymmetry grows with increasing virtual photon
rapidity. However, a virtual photon rapidity larger than 2 is not reachable at RHIC energy for the
kinematical region under consideration. Finally, the asymmetry at LHC energy plotted in Fig.~6
is similar but about a factor 2-3 smaller than that at RHIC.
\begin{figure}[hbtp]
\begin{center}
\includegraphics[angle=0,scale=1.1]{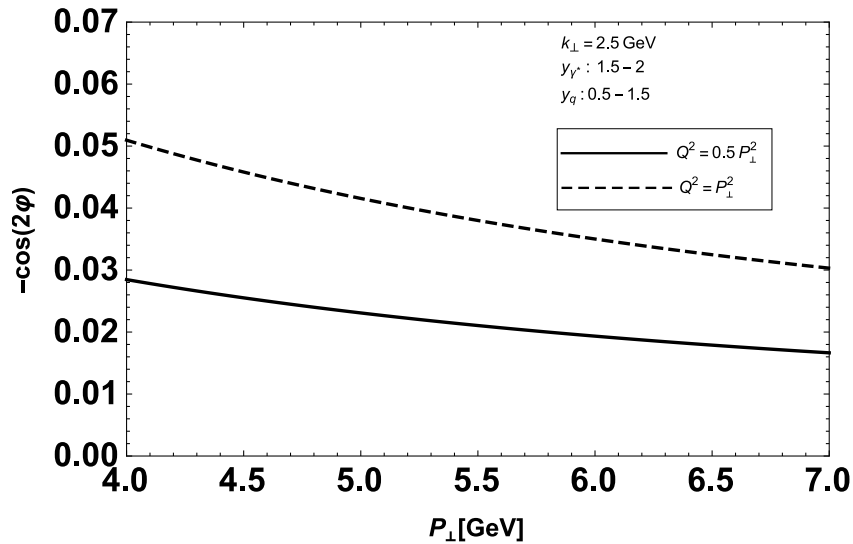}
\caption{Estimates of the azimuthal asymmetry as function of $P_{\perp}$ with $Q^2= P_\perp^2$ and
$Q^2= 0.5 P_\perp^2$,
 at $k_\perp=2.5 \ \text{GeV}$ and $\sqrt{s}$=200 GeV.
 The quark and virtual photon rapidities are integrated over the regions $y_q$[1.5, 2] and $y_{\gamma^*}$[0.5, 1.5].}
  \label{figs:v2 as Pperp}
\end{center}
\end{figure}
\begin{figure}[hbtp]
\begin{center}
\includegraphics[angle=0,scale=1.1]{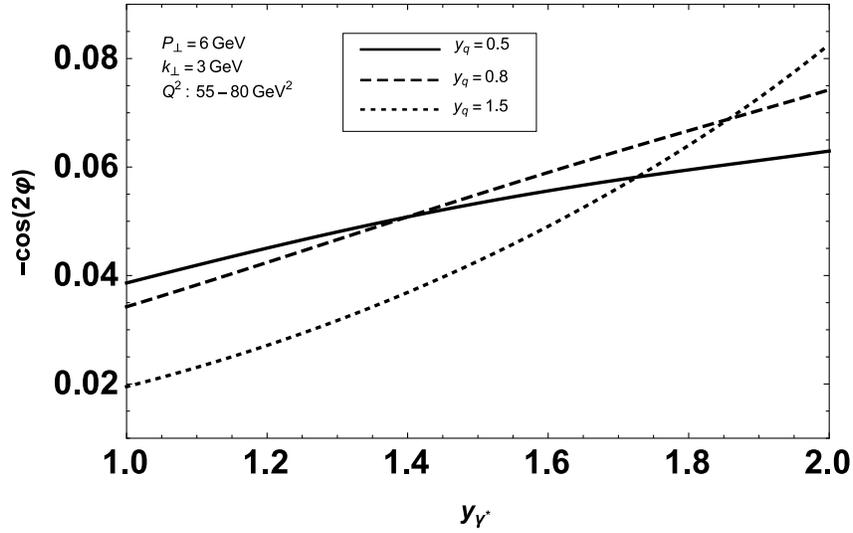}
\caption{The azimuthal asymmetry evaluated at $k_\perp=3 \, \text{GeV}$, $P_\perp=6 \, \text{GeV}$,
$\sqrt{s}$=200 GeV, and for the different quark rapidities $y_q=0.5, \ 0.8, \ 1.5$, respectively.
$Q^2$ is integrated over the region [55, 80] $\text{GeV}^2$} . \label{figs:v2 as y}
\end{center}
\end{figure}
\begin{figure}[hbtp]
\begin{center}
\includegraphics[angle=0,scale=1.1]{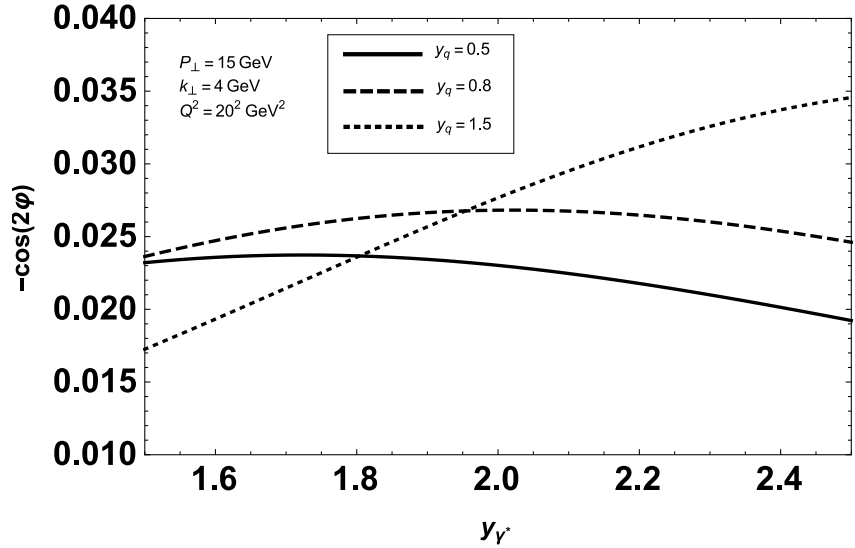}
\caption{The azimuthal asymmetry evaluated at $k_\perp=4 \, \text{GeV}$, $P_\perp=15 \,
\text{GeV}$, $Q = 20\ \text{GeV}$,  $\sqrt{s}$=5.02 TeV and for the different quark rapidities
$y_q=0.5, \ 0.8, \ 1.5$, respectively.} \label{figs:v2 as y at LHC}
\end{center}
\end{figure}

\section{Summary}
At small $x$ the linearly polarized gluon TMD is expected to be comparable in size to the
unpolarized gluon TMD, reflecting that the Color Glass Condensate can be highly polarized. In the
MV model, the linearly polarized gluon TMD saturates the positivity bound for the dipole case,
which means that it is in fact identical to the unpolarized gluon TMD. This relation persists under
small-$x$ evolution. Such a large effect is very promising for its experimental investigation,
especially since the linearly polarized gluon TMD has not been studied experimentally thus far.
From a theoretical point of view, the cleanest channel to probe the dipole type linearly polarized
gluon TMD is the $\cos 2\phi$ azimuthal asymmetry in virtual photon-jet pair production in $pA$
collisions, which can be studied at RHIC and LHC.

Despite that there is no reason to expect the asymmetry to be small at small $x$, given the maximal size of the linearly polarized gluon TMD, we find that
the effect of the linear gluon polarization is strongly suppressed due to TMD evolution effects. In this paper, we have presented
numerical estimations of the ratio between the linearly polarized gluon TMD and the unpolarized
gluon TMD and of the azimuthal asymmetry taking into account TMD evolution. The ratio after evolution
peaks at a transverse momentum on the order of the saturation
scale, where its maximal value for instance at scale $\mu^2=36 \ \text{GeV}^2$ is about 0.4 and at $\mu^2 = M_Z^2$ only on the few percent level.
Despite that the linear gluon TMD enters just once in the $\cos 2\phi$ asymmetry, the Sudakov suppression of the
asymmetry is much stronger than the ratio of TMDs would suggest.
For the typical kinematic regions accessible at RHIC, the
maximal size of the azimuthal asymmetry is found to be around 7\%. For such values experimental measurements at RHIC would still seem
feasible though. We note that these values do depend on the
input distributions we have chosen, but nevertheless we expect these results to give a realistic reflection of the amount of Sudakov suppression
for this observable. The situation for LHC is slightly worse, with 2-3\% asymmetries, but employing the small-$x$ model as a starting
point is more justified in this case. We have also found that the azimuthal asymmetry in jet-$Z/W$ production at
LHC is almost completely washed out by the Collins-Soper evolution effect.

In conclusion, the experimental study of this asymmetry at RHIC seems the most promising option
and, despite the strong Sudakov suppression, may allow to test the $k_t$-resummation formalism in
the small-$x$ regime and the theoretical expectation that the Color Glass Condensate state is in
fact polarized.

\begin{acknowledgments}
J. Zhou thanks Andreas Metz for suggesting to study the
observable numerically. J. Zhou has been supported by the National Science Foundations of China under Grant No.\ 11675093,
and by the Thousand Talents Plan for Young Professionals. Ya-jin Zhou has been supported by the
National Science Foundations of China under Grant No.\ 11375104 and No.\ 11675092.
This research has been partially supported by the EU ``Ideas'' program QWORK
(contract 320389).
\end{acknowledgments}

\end{document}